\documentclass[graybox]{svmult}

\usepackage{mathptmx}       
\usepackage{helvet}         
\usepackage{courier}        
\usepackage{type1cm}        
\usepackage{makeidx}         
\usepackage{graphicx}        
\usepackage{multicol}        
\usepackage[bottom]{footmisc}

\usepackage{lmodern}

\usepackage[T1]{fontenc}
\usepackage{geometry}
\usepackage{color}
\usepackage{amsmath}
\usepackage{amssymb}
\usepackage{cancel}
\usepackage{stmaryrd}
\usepackage{graphicx}
\usepackage{esint}
\usepackage{hyperref}
\usepackage{cite}
\usepackage{enumerate}

\newcommand{\ket}[1]{|{#1}\rangle}

\makeindex             

\begin{document}

\title*{Gauge Theories with Ultracold Atoms}
\author{Jo\~{a}o C. Pinto Barros, Michele Burrello and Andrea Trombettoni}
\institute{Jo\~{a}o C. Pinto Barros \at Albert Einstein Center for Fundamental Physics, Institute for Theoretical Physics, University of Bern, Sidlerstrasse 5, CH-3012 Bern, Switzerland, \email{jpintobarros@itp.unibe.ch}
\and Michele Burrello \at Center for Quantum Devices and Niels Bohr International Academy, Niels Bohr Institute,
University of Copenhagen, DK-2100 Copenhagen, Denmark, \email{michele.burrello@nbi.ku.dk}
\and Andrea Trombettoni \at 
CNR-IOM DEMOCRITOS Simulation Center and SISSA, Via Bonomea 265, I-34136 Trieste, Italy, \email{andreatr@sissa.it}}
%
%
\maketitle

\abstract*{We discuss and review in this chapter the developing field of 
research of quantum simulation of gauge theories with ultracold atoms.}

\abstract{We discuss and review in this chapter the developing field of 
research of quantum simulation of gauge theories with ultracold atoms.}

\section{Introduction}
\label{sec:1}
During the School on ``Strongly Coupled Field Theories for 
Condensed Matter and Quantum Information Theory'', held in Natal (Brazil) 
in the days 2 - 14 August 2015, one of the authors gave a course 
on ``Quantum Simulations of Gauge Fields with Ultracold Atoms''. The course was 
meant to be informal, at the blackboard, with time for discussions and 
to interact with the younger part of the audience. Subsequently, in 2017  
J. C. Pinto Barros obtained the PhD in SISSA (Trieste) defending a Thesis on 
``Field and Gauge Theories with Ultracold Atoms'', under the supervision 
of Andrea Trombettoni and Marcello Dalmonte and with Michele Burrello and 
Enrique Rico Ortega acting as external referees. The present chapter is based 
on the Natal's course and on the above mentioned PhD Thesis of 
J. C. Pinto Barros \cite{jpbthesis}. The latter Thesis is available at 
\verb|http://www.statphys.sissa.it/wordpress/?page_id=1095| 

The course gave an introductory discussion on lattice gauge theories, and then 
moved to explain how to simulate gauge potentials and gauge fields. A prior 
knowledge of ultracold atomic systems was assumed, even though during the 
lectures the corresponding concepts and notions were briefly reminded.

\section{Gauge theories\label{sec:Gauge-Theories}}

A gauge theory is a model that has a gauge symmetry. Such symmetry
can be seen as a redundancy in the description of the degrees of freedom. In other
words, this means that one can have two mathematically distinct solutions
of the equations describing the system and nonetheless they describe
the same physical situation. The paradigmatic example is classical electrodynamics. It describes the behavior of the
electric field $\vec{E}\left(t,\vec{x}\right)$ and the magnetic field
$\vec{B}\left(t,\vec{x}\right)$ in the presence
of an electric charge density $\rho\left(t,\vec{x}\right)$ and the
current density $\vec{j}\left(t,\vec{x}\right)$. The system is governed
by the Maxwell equations:
\begin{equation}
\begin{array}{cc}
\nabla\cdot\vec{E}\left(t,\vec{x}\right)=\rho\left(t,\vec{x}\right)\,;\qquad\hfill & \nabla\times\vec{B}\left(t,\vec{x}\right)-\partial_{t}\vec{E}\left(t,\vec{x}\right)=\vec{j}\left(t,\vec{x}\right)\,;\\
\nabla\cdot\vec{B}\left(t,\vec{x}\right)=0\,;\hfill & \nabla\times\vec{E}\left(t,\vec{x}\right)+\partial_{t}\vec{B}\left(t,\vec{x}\right)=0\,.\hfill
\end{array} \label{maxwell}
\end{equation}
In the above equations and in the rest of this Chapter, natural units
shall be adopted.
The homogeneous equations, which are independent of charges and currents,
can be straightforwardly solved by introducing a scalar potential $\phi\left(t,\vec{x}\right)$
and a vector potential $\vec{A}\left(t,\vec{x}\right)$:
\begin{equation}
\vec{E}\left(t,\vec{x}\right)=-\nabla\phi\left(t,\vec{x}\right)-\partial_{t}\vec{A}\left(t,\vec{x}\right)\,, \qquad \vec{B}\left(t,\vec{x}\right)=\nabla\times\vec{A}\left(t,\vec{x}\right)\,. \label{eq:EB_phiA}
\end{equation}
Using these two relations the last two equations in \eqref{maxwell} 
are fulfilled and the ones from the first row can be written in terms of
$\phi\left(t,\vec{x}\right)$ and $\vec{A}\left(t,\vec{x}\right)$.
After a solution is found, it can be plugged in Equation \ref{eq:EB_phiA}
in order to obtain the electric and magnetic fields. However, not all
different $\phi\left(t,\vec{x}\right)$ and $\vec{A}\left(t,\vec{x}\right)$
will give different electric and magnetic fields. In fact if two
fields $\phi\left(t,\vec{x}\right)^{\prime}$ and $\vec{A}\left(t,\vec{x}\right)^{\prime}$
are related to another solution $\phi\left(t,\vec{x}\right)$ and $\vec{A}\left(t,\vec{x}\right)$ by:
\begin{equation}
\phi\left(t,\vec{x}\right)^{\prime}=\phi\left(t,\vec{x}\right)+\partial_{t}\alpha\left(t,\vec{x}\right)\,, \qquad  \vec{A}\left(t,\vec{x}\right)^{\prime}=\vec{A}\left(t,\vec{x}\right)-\nabla\alpha\left(t,\vec{x}\right)\,,\label{eq:EMG_gauge}
\end{equation}
for some regular function $\alpha\left(t,\vec{x}\right)$, then the electric
and magnetic fields, given by Equation \ref{eq:EB_phiA}, remain unchanged.
This means that the solutions $\phi,\vec{A}$ and $\phi^{\prime},\vec{A}^{\prime}$
correspond to the same physical situation and therefore they are just
redundant descriptions of the same physics. The transformations of
Equation \ref{eq:EMG_gauge} are called gauge transformations.

The existence of a gauge symmetry does not require the field to be dynamical. 
Consider a charged quantum particle in a background
of a classical electromagnetic field. The Schr\"{o}dinger equation for
this system can be written as the equation in the absence of any field and 
``correcting'' the canonical momentum $\vec{p}\rightarrow\vec{p}-e\vec{A}$.
In the presence of an electromagnetic field the mechanical momentum,
associated with the kinetic energy of the particle and denoted here
by $\vec{\pi}$, is no longer the canonical momentum given by $\vec{p}$.
The relation between them is $\vec{\pi}=\vec{p}-e\vec{A}$ which is
at the core of this substitution. The same happens for the time derivative
with the scalar potential $i\partial_{t}\rightarrow i\partial_{t}-e\phi$.
Therefore, the Schr\"{o}dinger equation reads, in the absence of any other interactions:
\begin{equation}
\left(i\partial_{t}-e\phi\right)\psi\left(t,\vec{x}\right)=\left(-i\nabla-e\vec{A}\right)^{2}\psi\left(t,\vec{x}\right)\,.
\end{equation}
Also this equation is invariant under the transformation \ref{eq:EMG_gauge}
provided that the wave function is transformed by a position-dependent phase:
\begin{equation}
\psi\left(t,\vec{x}\right)=e^{ie\alpha\left(t,\vec{x}\right)}\psi\left(t,\vec{x}\right)\,.
\end{equation}
Given the space and time dependence of this transformation, it is denoted as 
a local gauge symmetry.

In quantum field theory an illustrative example is provided by QED.
The Lagrangian is given by:
\begin{equation}
{\cal L}=\bar{\psi}\left(\gamma^{\mu}\left(i\partial_{\mu}-eA_{\mu}\right)-m\right)\psi-\frac{1}{4}F_{\mu\nu}F^{\mu\nu}\,.\label{eq:WED_Lagrangian}
\end{equation}
Implicit sum over repeated indices is assumed. $\gamma^{\mu}$ are the gamma matrices satisfying
the Clifford algebra $\left\{ \gamma^{\mu},\gamma^{\nu}\right\} =2\eta^{\mu\nu}$,
$\eta^{\mu\nu}$ is the Minkowski metric $\eta=\mathrm{Diag}\left(1,-1,-1,-1\right)$,
$\psi$ the Dirac spinor and $\bar{\psi}=\psi^{\dagger}\gamma^{0}$.
The indices $\mu$ run from $0$ to $3$ where $0$ corresponds to
the time index. $A_{\mu}$ is called gauge field and the last
term of the Lagrangian corresponds to its kinetic term where $F_{\mu\nu}=\partial_{\mu}A_{\nu}-\partial_{\nu}A_{\mu}$.
Also in this case there is a local set of transformations that leave this
Lagrangian invariant. Explicitly:
\begin{equation}
A_{\mu}\left(x\right)\rightarrow A_{\mu}\left(x\right)-\frac{1}{e}\partial_{\mu}\alpha\left(x\right)\,, \qquad \psi\left(x\right)\rightarrow e^{i\alpha\left(x\right)}\psi\left(x\right)\,,\label{eq:U1QED}
\end{equation}

One can define the covariant derivative $D_{\mu}=\partial_{\mu}+ieA_{\mu}^{a}$ such that, under a gauge transformation, $D_{\mu}\psi \rightarrow e^{i\alpha\left(x \right)} D_{\mu}\psi$. In this way, the local gauge symmetry becomes apparent.

This is an example of a $U\left(1\right)$ gauge theory: a gauge transformation
is defined, at each point, by phases $\alpha\in\left[0,2\pi\right[$
which combine according to the group $U\left(1\right)$. This construction can be generalized to other gauge groups, like $\mathbb{Z}_{N}$,
or even non-Abelian, like $SU\left(N\right)$, for $N$ an integer
number. For example, the Kitaev toric code is a $\mathbb{Z}_{2}$ (Abelian)
gauge theory \cite{kitaev2003} whereas Quantum Chromodynamics (QCD),
the theory that describes strong interactions in particle physics,
is a $SU\left(3\right)$ (non-Abelian) gauge theory \cite{gross1973,fritzsch1973,politzer1973}.
In the following a brief description of non-Abelian $SU\left(N\right)$ gauge invariance
in quantum field theory is provided. For more details see, for example, \cite{peskin1995}.

In order to explore these other symmetries, 
extra indices must be inserted (in the paradigmatic example of QCD
these are the color indices). To simplify the notation, whenever
$\psi$ it is used it is meant:
\begin{equation}
\psi\equiv\left(\begin{array}{c}
\psi_{1}\\
\psi_{2}\\
\vdots\\
\psi_{n}
\end{array}\right)\label{eq:psi_color}
\end{equation}
where each one of the $\psi_{i}$ corresponds to a (four-component
in $3+1$ dimensions) Dirac spinor. Consider then a general symmetry
group and a respective set of generators represented by Hermitian
$n\times n$ matrices $t^{a}$. The goal is to build a Lagrangian
which is invariant under the set of local transformations 
\begin{equation}
\psi\left(x\right)\rightarrow e^{i\alpha^{a}\left(x\right)t^{a}}\psi\left(x\right)\,.
\end{equation}
This is a unitary transformation that mixes the $n$ components of the vector \ref{eq:psi_color} following a $n-$dimensional representation of the gauge group element $e^{i\alpha^{a}\left(x\right)t^{a}}$. The gauge field becomes in turn a matrix which can be parametrized as $A_\mu^a t^a$. Under a gauge transformation the field transforms as
\begin{equation}
A_\mu^a t^a\rightarrow e^{i\alpha^{a}\left(x\right)t^{a}}\left(A_\mu^a t^a+\frac{i}{g}\partial_\mu\right)e^{-i\alpha^{a}\left(x\right)t^{a}}.
\end{equation}
Writing the covariant derivative as $D_{\mu}=\partial_{\mu}-igA_{\mu}^{a}t^{a}$ one finds $D_{\mu}\psi\left(x\right)\rightarrow e^{i\alpha^{a}\left(x\right)t^{a}}D_{\mu}\psi\left(x\right)$. In this way $\bar{\psi}\left(\gamma^{\mu}D_{\mu}-m\right)\psi$ is a gauge invariant operator which includes the fermionic kinetic term and the matter-gauge coupling. Note that $\bar{\psi}$ is to be interpreted as line vector
with components $\bar{\psi}_{i}$ and $\gamma^{\mu}$ are diagonal
on the color indices, i.e. act the same for every color by standard
matrix multiplication $\gamma^{\mu}\psi_{i}$.
In order to define the gauge field dynamics, its gauge invariant kinetic term must be inserted. A possible way to derive its form is by considering the commutator $\left[D_\mu,D_\nu\right]=it^a F_{\mu\nu}^a$. Putting it differently, a general form for $F_{\mu\nu}^a$ can be obtained from this formula. Explicit computation yields $F_{\mu\nu}^{a}=\partial_{\mu}A_{\nu}^{a}-\partial_{\nu}A_{\mu}^{a}+gf^{abc}A_{\mu}^{b}A_{\nu}^{c}$ where the structure constants $f^{abc}$ are given by $\left[t^{a},t^{b}\right]=it^{c}f^{abc}$ and depend only on the symmetry group. From the transformation law for the covariant derivatives, one can see that $F_{\mu\nu}^{a}F^{a\mu\nu}$ is gauge invariant. The full Lagrangian can then be written as
\begin{equation}
{\cal L}=\bar{\psi}\left(\gamma^{\mu}D_{\mu}-m\right)\psi-\frac{1}{4}F_{\mu\nu}^{a}F^{a\mu\nu}\label{L_matter-gauge}
\end{equation}

The perspective of implementing these kind of models in table top experiments is very appealing. First of all, it could give answers to very fundamental questions in physics like, for example, the exploration of the phase diagram of QCD. That is certainly a long term challenge and the path envisioned towards it involves the implementation of simpler intermediate steps. While QCD has a $SU\left(3\right)$ gauge symmetry and involves $3+1$ dimensions, this does not need to be the main target. A much simpler case of a $U\left(1\right)$ gauge symmetry in $1+1$ dimensions is already of great interest. In fact, this was the target of the first experimental implementation of a lattice gauge theory \cite{martinez2016}(to be discussed in Section \ref{subsec:Encoding-in-}). Step by step one may 
think to be able to realize more and more complex models. It is clear 
that if these models are realized they become interesting 
on their own both theoretically and experimentally. 
In particular, for example, it may also be advantageous to have situations where only certain degrees of freedom live in higher dimensions keeping others in lower dimensionality \cite{amaral1992,PintoBarros2017LR,PintoBarros1017Robustness} which could be used 
to simulate systems with long-range interactions, which have been the subject 
of an intense investigation in the last years 
\cite{PhysRevE.89.062120,Brezin2014,PhysRevE.92.052113,PhysRevB.94.224411,PAULOS2016246,PhysRevE.95.012143,PhysRevE.96.012108,PhysRevLett.118.241601,Behan_2017,PhysRevB.96.104432}. 

\subsection{Gauge symmetry on the lattice}

\subsubsection{Static fields \label{sec:static}}

Following the discussion in the previous Section, a many-body Hamiltonian in the presence of a magnetic field can be obtained by replacing the momentum components for each particle by $p_{i}\rightarrow p_{i}-eA_{i}$. On the lattice, instead, this can be approximated by the Peierls substitution where the hopping parameters become complex. This is valid in a tight-binding regime and for a slow varying magnetic
field. Explicitly the kinetic term is modified according to
\begin{equation}
K=\underset{\vec{r},j}{\sum}t_{j}\hat{a}_{\vec{r}+\hat{j}}^{\dagger}\hat{a}_{\vec{r}}+\mathrm{h.c.}\rightarrow\underset{\vec{r},j}{\sum}t_{j}\hat{a}_{\vec{r}+\hat{j}}^{\dagger}e^{i\theta_{j}\left(\vec{r}\right)}\hat{a}_{\vec{r}}+\mathrm{h.c.}
\end{equation}
In the previous equation the sum of $\vec{r}$ is taken over the lattice sites
and the sum of $j$ is taken over all $d$ directions corresponding
to the dimensionality of the system. The angle $\theta_{j}\left(\vec{r}\right)$
is just a phase that can depend, on general grounds, on both the direction
of the hopping and the position.
The key difference is that this phase here is non dynamical, so there
is no kinetic term for it. This simply corresponds to allow for the hopping parameter of the particles on the lattice
to be complex. Similarly to the models in continuum space, not all complex hoppings represent different physical
scenarios as there is gauge invariance. In Section \ref{sim_gauge_pot}, several examples
of techniques to engineer complex phases on the hopping parameters
are discussed. Reviews can be found in \cite{dalibard2011,Goldman2014b,dalibard2015,burrello2017}. 

\subsubsection{Dynamical fields \label{sec:dynamical}}

In order to study a dynamical quantum (lattice) gauge theory, the lattice system under analysis must include also the degrees of freedom for the gauge fields and the complex hopping parameters are therefore promoted to operators acting on these degrees of freedom. Such degrees of freedom are usually associated to the lattice edges and their kinetic term must be supplied. 

A constructive way to define such a system consists on taking the Lagrangian \ref{L_matter-gauge}, write an Hamiltonian and perform a naive discretization. This offers in turn a recipe to engineer possible quantum simulations of these systems: a straightforward way of proceeding is indeed to create a system implementing the specific Hamiltonian of the lattice gauge theory, therefore it is useful to consider such theories in their Hamiltonian formulation \cite{kogut1975hamiltonian}. 

To describe the $U\left(1\right)$ case, it is useful to introduce the following link operators, acting on the gauge degrees of freedom:
\begin{equation}
U_{\vec{r},j}=\exp\left(ie\overset{a\left(\vec{r}+\hat{j}\right)}{\underset{a\vec{r}}{\int}}dxA_{j}\left(x\right)\right)\,,\qquad L_{\vec{r},j}=\frac{E_{\vec{r},j}}{e}\,.
\end{equation}
$U$ and $L$ are operators corresponding respectively to the connection and electric field of the theory (see, for example, \cite{ZoharBurrello2015} for more details). Based on these operators, we can define the Hamiltonian
\begin{equation}
H=-\frac{i}{2a}\underset{\vec{r},j}{\sum}\left(\psi_{\vec{r}}^{\dagger}U_{\vec{r},j}\psi_{\vec{r}+\hat{j}}-\mathrm{h.c.}\right)+m\underset{n}{\sum}\psi_{\vec{r}}^{\dagger}\psi_{\vec{r}}+\frac{ae^{2}}{2}\underset{n}{\sum}L_{\vec{r},j}^{2}\,,
\end{equation}
which reproduces the correct continuum theory when the \textit{naive} continuum limit is taken. In the expressions above $a$ is the lattice spacing, $\vec{r}$ are the lattice points and $j$ labels the links connected to it. $E_{\vec{r}j}$ is the discretized version of the electric field which is the conjugate momentum of $A_j$ in the continuum version. The commutation relations between the link operators are
\begin{equation}
\left[L_{\vec{r},i},U_{\vec{r}^{\prime},j}\right]=\delta_{\vec{r}\vec{r}^{\prime}}\delta_{ij}U_{\vec{r},i},\quad \left[L_{\vec{r},i},U_{\vec{r}^{\prime},j}^{\dagger}\right]=-\delta_{\vec{r}\vec{r}^{\prime}}\delta_{ij}U_{\vec{r},i}^\dagger,\quad
\left[U_{\vec{r},i},U_{\vec{r}^{\prime},j}\right]=\left[U^\dag_{\vec{r},i},U_{\vec{r}^{\prime},j}\right]=0.
\label{eq:com-rel}
\end{equation}
We pause here to point out a couple of subtleties. The first concerns the so-called \textit{naive} continuum limit, obtained by simply sending $a\rightarrow0$. While this works well for bosons, fermions suffer from the so-called "fermion doubling problem". When this limit is taken with more care each fermion flavor on the lattice gives rise to $2^{d}$ fermion flavors on the continuum, being $d$ the number of discretized dimensions. The Nielsen-Ninomiya
Theorem \cite{Nielsen1981,Nielsen1981I,Nielsen1981II} states that this is always the case when the fermion action is real, local and invariant under lattice translations and chiral transformations. 
There are alternative approaches to evade the Nielsen-Ninomiya Theorem which have their own advantages and disadvantages. A possible choice, popular among the quantum simulation community, is provided by staggered fermions \cite{Susskind1977} (also known as Kogut-Susskind fermions). The idea consists on distributing the spinor components among different lattice sites. In this way, instead of a spinor per site, one has only one fermion. Only for the Hamiltonian formulation of the $1+1D$ theory the fermion doubling problem can be completely solved in this way. In this case the Hamiltonian becomes
\begin{equation}
H=-\frac{i}{2a}\underset{n}{\sum}\left(c_{n}^{\dagger}U_{n}c_{n+1}-\mathrm{h.c.}\right)+m\underset{n}{\sum}\left(-1\right)^{n}c_{n}^{\dagger}c_{n}+\frac{ae^{2}}{2}\underset{n}{\sum}L_{n}^{2}.
\end{equation}
Spinors can be reconstructed from $\psi_{n}=\left(c_{2n}, c_{2n+1}\right)^T{/\sqrt{a_{st}}}$.

In higher dimensions the most non-trivial step consists on the existence of plaquette terms or, in other words, an energy cost for magnetic fields. These are gauge-invariant terms which must be present in order to fully represent the gauge theory. The absence of these terms is related to the strong coupling limit of the theory. On a 2D square lattice, the plaquette term originating at the point $\vec{r}$ is $U_{\square}=U_{\vec{r},x}U_{\vec{r}+\hat{x},y}U_{\vec{r}+\hat{y},x}^{\dagger}U_{\vec{r},y}^{\dagger}$ consisting on the smallest loops possible to draw on the lattice. The Hamiltonian for $d$ spatial dimensions takes the form:
\begin{equation}
\begin{split}
H=&-\frac{i}{2a}\underset{\vec{r},i}{\sum}\left(-1\right)^{r_{1}+\ldots+r_{i-1}}\left(c_{\vec{r}}^{\dagger}U_{\vec{r},i}c_{\vec{r}+\hat{i}}-\mathrm{h.c.}\right)+m\underset{\vec{r}}{\sum}\left(-1\right)^{r_{1}+\ldots+r_{d}}c_{\vec{r}}^{\dagger}c_{\vec{r}}\\
&+\frac{a^{2-d}e^{2}}{2}\underset{\vec{r},i}{\sum}L_{\vec{r},i}^{2}-\frac{a^{d-4}}{4e^{2}}\underset{\square}{\sum}\left(U_{\square}+U_{\square}^{\dagger}\right)\label{eq:H_KG_general_d}
\end{split}
\end{equation}
The extra alternating signs on the first term are required to obtain the correct Dirac Hamiltonian in the continuum limit with staggered fermions \cite{Susskind1977}.

Another fundamental point, associated with the Hamiltonian formulation, consists on the restriction of the Hilbert space to physical states. This can be derived from the Lagrangian formulation by noting that the component $A_{0}$ is non-dynamical (there is no term $\partial_{0}A_{0}$).
As a consequence it acts as a Lagrange multiplier enforcing the Gauss' law as a constraint. Therefore, on the one spatial dimensional lattice, the physical states are defined by the relation:
\begin{equation} \label{gausslaw}
G_{\vec{r}}\left|\Psi\right\rangle =0
\end{equation}
for each lattice site $\vec{r}$, where
\begin{equation}
G_{\vec{r}}=\underset{i}{\sum}\left(L_{\vec{r},j}-L_{\vec{r}-\hat{j},j}\right)-Q_{\vec{r}}\,, \label{eq:u1-generators}
\end{equation}
where $Q_{\vec{r}}$ is the dynamical matter charge. For the $1+1D$ case, for example, $Q_{n}=c_{n}^{\dagger}c_{n}+\frac{1-\left(-1\right)^{n}}{2}$. The alternating tem, which may look odd, is related to the staggered formulation. Considering a state with no electric field. The Gauss law demands that fermions
populate the odd sites while leaving the even empty.
This is because the
spinor degrees of freedom are distributed along the lattice. Occupied
odd sites have the interpretation of a filled Dirac sea. When a fermion
hops from an odd to an even site it creates a hole in the Dirac sea
while creating a particle above the Dirac sea. This is
interpreted as the creation of particle/anti-particle pair where the
hole plays the role of an anti-particle. In the presence of gauge
fields, the hopping described above must be accompanied by a change on
the electric field preserving Gauss' law, as described by the connection operator in the first term of the Hamiltonian \eqref{eq:H_KG_general_d}. 

The $G_{\vec{r}}$ are also generators of the gauge transformation and can be extended for the $U\left(N\right)$ and $SU\left(N\right)$ gauge theories. 
To this purpose, one can consider matter fields $\psi_{\vec{r}}$ that transform under the gauge symmetries under a suitable representation of dimension $n$ of the gauge group. The generators of the gauge symmetries must therefore satisfy the relation $\left[G_{\vec{r}}^{a},\psi_{\vec{r}}\right]=t^{a}\psi_{\vec{r}}$ where $t^{a}$ are $n-$dimensional representations of the (left) group generators.

In order to preserve the gauge-invariance of the Hamiltonian, the connection operators must transform like tensors under the gauge transformations and they must follow the same representation of the matter fields: $U_{\vec{r},{j}} \to e^{i\alpha_{\vec{r}}^a t^a} U_{\vec{r},{j}} e^{-i\alpha_{\vec{r}+\hat{j}}^at^a}$. In particular the connection is multiplied on the left side by the transformation inherited from the lattice site on its left and on the right side by the inverse of the transformation inherited from the lattice site on its right.
When we deal with a non-Abelian group, it is thus useful to distinguish left and right generators for the group transformations \cite{kogut1975hamiltonian}, labelled by $L_{\vec{r},i}$ and $R_{\vec{r},i}$ respectively (see, for example, \cite{ZoharBurrello2015}). 
The local generators of the gauge transformation can therefore be defined as:
\begin{equation}
G_{\vec{r}}^{a}=\sum_{i}\left(L_{\vec{r},i}^{a}+R_{\vec{r}-\hat{i},i}^{a}\right)+\psi_{\vec{r}}^{\dagger}t^{a}\psi_{\vec{r}}.
\end{equation}
Finally, the lattice Hamiltonian for the non-Abelian theory will be:
\begin{multline}
H=-\frac{i}{2a}\underset{\vec{r},i}{\sum}\left(-1\right)^{r_{1}+\ldots+r_{i-1}}\left(\psi_{\vec{r}}^{\dagger}U_{\vec{r}i}\psi_{\vec{r}+\hat{i}}-\mathrm{h.c.}\right)+m\underset{\vec{r}}{\sum}\left(-1\right)^{r_{1}+\ldots+r_{d}}\psi_{\vec{r}}^{\dagger}\psi_{\vec{r}} \\
+\frac{a^{2-d}g^{2}}{2}\underset{\vec{r},i,a}{\sum}\left(\left(L_{\vec{r}i}^{a}\right)^{2}+\left(R_{\vec{r}i}^{a}\right)^{2}\right)-\frac{a^{d-4}}{4g^{2}}\underset{\square}{\sum}\mathrm{Tr}\left(U_{\square}+U_{\square}^{\dagger}\right)
.\label{eq:General_Abelian_QL}
\end{multline}
Again, the Gauss law should be imposed on physical states $G_{\vec{r}}^a\left|\Psi\right\rangle =0$. 

Often times, in the proceeding Sections, the matter-gauge correlated hopping will be written as $\psi_{\vec{r}}^{\dagger}U_{\vec{r}i}\psi_{\vec{r}+\hat{i}}+\mathrm{h.c.}$ rather than $i\left(\psi_{\vec{r}}^{\dagger}U_{\vec{r}i}\psi_{\vec{r}+\hat{i}}-\mathrm{h.c.}\right)$ as above. While the latter reproduces the familiar continuum Hamiltonian in the naive continuum limit, both are related by a gauge transformation.


\subsubsection{Challenges, limitations and quantum link models\label{sec:QL}}

Cold atom systems offer the possibility to construct Hubbard-like Hamiltonians with tunable hopping parameters and on-site interactions. However, gauge potentials and gauge fields demand more than that.

When the field is static, as described in Section \ref{sec:static}, the hopping parameters become complex. This is not readily available in simple optical lattices, but, thanks to recent experimental developments, it is nowadays possible to engineer static gauge fields, as we will discuss in Section \ref{sim_gauge_pot}.

For dynamical gauge fields, as discussed in Section \ref{sec:dynamical}, the matter hopping and link operators must be correlated in such a way to guarantee the existence of the local gauge symmetry (at each lattice site). Such kind of hopping is not natural in a cold atomic system and a discussion on how to implement is presented in Section \ref{sec:Simulating-gauge-fields}.

There is, yet, a further difficulty for dynamical gauge fields. Take, for example, the commutation relations in \ref{eq:com-rel} pertaining a certain link. The operator $U_{\vec{r}}$ acts as a raising operator of the electric field (or equivalently of $L_{\vec{r}}$). But, for a $U(1)$ theory, this corresponds to an infinite Hilbert space per link. Constructing such links is certainly a challenge for its implementation, even for small lattice sizes. A solution of this problem is provded by quantum link models which are characterized by a finite Hilbert space per link, without violating the required gauge symmetry. These models were introduced by Horn in 1981 \cite{horn1981} and were further studied in \cite{orland1990,orland1992,chandrasekharan1997,brower1999,brower2004}.
Proposed as an alternative formulation to Wilson gauge field theories on the lattice, they became an attractive realization of gauge symmetries for quantum simulation purposes. 

In quantum link models the link degrees of freedom are replaced by quantum spins, such that the algebra in \ref{eq:com-rel} is replaced by the algebra of angular momentum. In particular this correspond to considering alterative link operators:
\begin{equation}
L_{+\vec{r},i}=S_{\vec{r},i}^{x}+iS_{\vec{r},i}^{y}\,,\quad L_{\vec{r},i}=S_{\vec{r},i}^{z}
\end{equation}
where the raising and lowering operators $L_{\pm \vec{r},i}$ replace $U_{\vec{r}}$ and its conjugate.
With this construction, the first two relations of \ref{eq:com-rel} are still satisfied. However, the last no longer holds because $U$ and $U^\dag$ do not commute any longer. In particular, $L_{\pm \vec{r}i}$ are not unitary whereas $U_{\vec{r}}$ was.

Even though the algebra itself is different, the angular momentum operators can be
equally used to construct a gauge theory without compromising the
gauge symmetry. In particular, we can choose the dimension of the Hilbert space in each of the
links to be $2S+1$ with $S$ a positive half integer (corresponding, in the spin language, to the total spin). It is expected that in
the limit of large $S$ the Wilson formulation should be recovered.
Explicitly one can use the following link variables $U_{\vec{r}i}\rightarrow L_{+\vec{r}i}/\sqrt{S\left(S+1\right)}$.
The new non-zero commutation relation is
$\left[U_{\vec{r}i},U_{\vec{r}i}^{\dagger}\right]=2L_{\vec{r}i}/S\left(S+1\right)$.
In the limit of $S\rightarrow+\infty$ the right hand side goes to zero and the initial algebra is recovered.


There is an analogous construction for $U\left(N\right)$ non-Abelian symmetries. One can see that the symmetry can be realized using an $SU\left(2N\right)$ algebra (note that for $N=1$ this gives, correctly, $SU\left(2\right)$). It is possible to construct the new algebra using  the so-called ``rishon fermions'' \cite{brower2004}. They are written in terms of pairs of fermionic operators $l_{\vec{r},j}^{m}$ and $r_{\vec{r}+\hat{j},j}^{m}$ for each link between the sites $\vec{r}$ and $\vec{r} + \hat{j}$. These operators define additional left and right gauge modes laying on the lattice sites $\vec{r}$ and $\vec{r} + \hat{j}$, with the aim of describing the link degrees of freedom. $m$ labels their color index. We can write:
\begin{align}
&L_{\vec{r},j}^{a}=\frac{1}{2}l_{\vec{r},j}^{m\dagger}t_{mn}^{a}l_{\vec{r},j}^{n}\,,\qquad R_{\vec{r},j}^{a}=\frac{1}{2}r_{\vec{r}+\hat{j},j}^{m\dagger}t_{mn}^{a}r_{\vec{r}+\hat{j},j}^{n}\,,\\&
E_{\vec{r},j}=\frac{1}{2}\left(r_{\vec{r}+\hat{j},j}^{m\dagger}r_{\vec{r}+\hat{j},j}^{m}-l_{\vec{r},j}^{m\dagger}l_{\vec{r},j}^{m}\right)\,,\\
&U_{\vec{r},j}^{mn}=l_{\vec{r},j}^{m}r_{\vec{r}+\hat{j},j}^{n\dagger}\,.\label{eq:rishon_representation}
\end{align}
The finiteness of the Hilbert space is a feature desirable for future quantum simulation schemes. Even though not a primary concern at this stage, it is reassuring that the effective continuum limit can be achieved even if one uses quantum link models \cite{chandrasekharan1997}.

\section{Simulation of gauge potentials \label{sim_gauge_pot}}

In accordance with the previous discussion, the goal of this Section is to show specific examples on how a complex hopping parameter can be engineered. The two main strategies described will be two contrasting situations. In one external parameters are varied adibatically (Section \ref{sec:adiabatic}), while in the other fast modes are integrated out (Section \ref{sec:driven}).

\subsection{Adiabatic change of external parameters \label{sec:adiabatic}}

The idea of this approach has, in its core, the tight relation between
the Aharonov-Bohm phase \cite{aharonov1959} and the Berry phase which
was a concept introduced by Berry in \cite{berry1984}. The first
is the phase acquired by a particle traveling around a closed
contour. At the end of the path, when it is back to the initial position,
the wave function acquires a new phase which is independent of the
details of how the path was done and only depending on the total magnetic
flux through the contour. On the other side, the Berry phase corresponds
to the phase acquired when some external parameters of the system
are varied in time, ``slowly'', coming back again to their initial
value for a non-degenerate state. In a more precise way, the starting
point is an Hamiltonian $H\left(q^{a},\lambda_{i}\right)$ where $q^{a}$
are degrees of freedom and $\lambda_{i}$ are a set of external parameters.
If these parameters are varied sufficiently slowly returning, in the
end, to their initial value, and if the initial state is an eigenstate non degenerate in energy, then
the system will be back to its initial state. During the process, however, it will acquire a phase:
\begin{equation}
\left|\psi\right\rangle \underset{\mathrm{adiabatic\ change}}{\longrightarrow}e^{i\gamma}\left|\psi\right\rangle \,.
\end{equation}
The phase $\gamma$ can be derived by computing the time evolution operator
and subtract the ``trivial'' dynamical phase acquired simply due to the time evolution. 
Let us consider the adiabatic evolution of a system such that each energy eigenstates remain non-degenerate during the whole process. In this case, starting from one of the eigenstates of the initial Hamiltonian, the system will continuosly evolve remaining in the corresponding eigenstate, with energy $E(t)$, at each time. Therefore the dynamical phase results $e^{-i\int E\left(t\right)dt}$. The additional Berry phase reads instead: 
\begin{equation}
\gamma=\ointctrclockwise_{{\cal C}}\tilde{A}_{i}\left(\lambda\right)d\lambda_{i}\,,
\end{equation}
where ${\cal C}$ is the closed path in the space of the parameters $\lambda_{i}$
and $\tilde{A}_{i}$ are given by:
\begin{equation}
\tilde{A}_{i}\left(\lambda\right)=i\left\langle \phi\left(\lambda\right)\right|\frac{\partial}{\partial\lambda_{i}}\left|\phi\left(\lambda\right)\right\rangle\,,  \label{eq:artificial_A}
\end{equation}
where $\left|\phi\left(\lambda\right)\right\rangle $ are reference
eigenstates taken with an arbitrary choice of their overall phases. 
$\tilde{A}\left(\lambda\right)$
is the Berry connection. Different choices of the reference eigenstates
with different phases, for example $\ket{\phi'(\lambda)} = e^{i\alpha\left(\lambda\right)}\left|\phi\left(\lambda\right)\right\rangle $,
would just reproduce a gauge transformation on $\tilde{A}$:
\begin{equation}
\tilde{A}_i\rightarrow\tilde{A}_i' = \tilde{A}_i-\frac{\partial\alpha}{\partial\lambda_{i}}\,.
\end{equation}
This principle can be applied in multi-level atomic systems in order
to reproduce artificial gauge fields in an ultracold atomic setting.
As an example, the computation can be done for a two level atom,
where it is shown how this vector potential appears explicitly at
the Hamiltonian level. These two levels correspond to two internal
states of the atom, the ground state $\left|g\right\rangle $ and an 
excited state
$\left|e\right\rangle $. The center of mass Hamiltonian, assumed
diagonal on the internal states, is taken to be simply the
free particle Hamiltonian. The total Hamiltonian is $H=H_{0}+U$. By
a suitable shift of the energy spectrum we
can assume that the ground and excited state energies are related by $E_{g}=-E_{e}$. Then $U$ can be written as
\begin{equation}
U=\frac{\Omega}{2}\left(\begin{array}{cc}
\cos\theta & e^{i\phi}\sin\theta\\
e^{i\phi}\sin\theta & -\cos\theta
\end{array}\right)
\end{equation}
where $\theta$ and $\phi$ may depend on the position. The frequency
$\Omega$ characterizes the strength of the coupling between the two
states and it is assumed to be position independent. The eigenstates
of this operator, denoted to as ``dressed states'', are given
by:
\begin{equation}
\left|\chi_{1}\right\rangle =\left(\begin{array}{c}
\cos\frac{\theta}{2}\\
e^{i\phi}\sin\frac{\theta}{2}
\end{array}\right)\,,\qquad \left|\chi_{2}\right\rangle =\left(\begin{array}{c}
-e^{-i\phi}\sin\frac{\theta}{2}\\
\cos\frac{\theta}{2}
\end{array}\right)
\end{equation}
with eigenvalues $\pm\hbar\Omega/2$ respectively. We assume
that the initial internal state is $\left|\chi_{1}\right\rangle $ and
that the evolution is adiabatic, 
such that the system remains in the state $\ket{\chi_1}$ at
all times. Hence the state of the system can be described by a wave
function $\left|\psi\left(t,\vec{r}\right)\right\rangle =\varphi\left(t,\vec{r}\right)\left|\chi_{1}\left(\vec{r}\right)\right\rangle $
where $\varphi\left(t,\vec{r}\right)$ will obey a modified Schr\"{o}dinger
equation due to the dependence of $\left|\chi_{1}\left(\vec{r}\right)\right\rangle $
on the position. Plugging this into the Schr\"{o}dinger equation and projecting
on $\left|\chi_{1}\left(\vec{r}\right)\right\rangle $, we find
the effective Hamiltonian governing $\varphi$:
\begin{equation}
H_{\mathrm{eff}}=\frac{\left(p_{i}-i\left\langle \chi_{1}\left(\vec{r}\right)\right|\frac{\partial}{\partial x_{i}}\left|\chi_{1}\left(\vec{r}\right)\right\rangle \right)^{2}}{2m}+\frac{\left|\left\langle \chi_{2}\left(\vec{r}\right)\right|\frac{\partial}{\partial x_{i}}\left|\chi_{1}\left(\vec{r}\right)\right\rangle\right| ^{2}}{2m}+\frac{\Omega}{2}
\end{equation}
As expected, a vector potential $\tilde{A}_{i}\left(\vec{r}\right)$ corresponding to the Berry connection is found.
Additionally a potential $\tilde{V}\left(\vec{r}\right)$ is also created
and related to virtual transitions to the other state $\left|\chi_{2}\left(\vec{r}\right)\right\rangle $.
In this two level approximation, these two quantities are given by
$\tilde{A}_{i}\left(\vec{r}\right)=\frac{\cos\theta-1}{2}\frac{\partial\phi}{\partial x_{i}}$
and $\tilde{V}\left(\vec{r}\right)=\frac{\left(\nabla\theta\right)^{2}+\sin^{2}\theta\left(\nabla\phi\right)^{2}}{8m}$\@.
Discussions about the practical implementation on optical lattices can be
found in \cite{dalibard2011,dum1996,visser1998,Goldman2014b}. First experimental
evidence of scalar potentials in quantum optics was found in \cite{Dutta1999}
and the first observation of geometric magnetic fields in cold atomic
physics was done in \cite{Lin2009}. By considering a set of degenerate
or quasi-degenerate dressed states it is possible to achieve non-Abelian
gauge potentials as well \cite{dalibard2011,Goldman2014b}. 

\subsection{Effective Hamiltonian in periodic driven system \label{sec:driven}}

In contrast to the approach of the previous Subsection, where the
creation of the magnetic field relied on a slow change in time 
(i.e. the particle moves slowly 
enough such that the position dependent internal state is followed 
adiabatically), the following technique relies on fast oscillations.
The basic principle consists on having two very distinct timescales.
A fast oscillating time dependent potential will give rise to an effective
time independent Hamiltonian which will present the desired complex
hopping term. A general technique was proposed in \cite{goldman2014}
and it is based on a generic time-dependent periodic Hamiltonian:
\begin{equation}
H=H_{0}+V\left(t\right)
\end{equation}
where all the the time dependence is relegated to $V\left(t\right)=V(t+\tau)$ where $\tau$ is the time period. $V(t)$ can be decomposed as:
\begin{equation}
V\left(t\right)=\underset{n}{\sum}\left(V_{n+}e^{in\omega t}+V_{n-}e^{in\omega t}\right)
\end{equation}
where $V_{n\pm}$ are operators and $\omega= 2\pi/\tau$. The condition $V_{n+}=V_{n-}^{\dagger}$ guarantees
the Hermiticity of the Hamiltonian. 

A unitary transformation $e^{iK\left(t\right)}$ generates an effective Hamiltonian given by:
\begin{equation}
H_{\mathrm{eff}}=e^{iK\left(t\right)}He^{-iK\left(t\right)}+i\left(\frac{\partial}{\partial t}e^{iK\left(t\right)}\right)e^{-iK\left(t\right)}
\end{equation}
We choose a periodic operator $K\left(t\right)$ such that the effective Hamiltonian is time independent.
Under this requirement, the time evolution operator
can be represented as:
\begin{equation}
U\left(t_{i}\rightarrow t_{f}\right)=e^{iK\left(t_{f}\right)}e^{-iH_{\mathrm{eff}}\left(t_{f}-t_{i}\right)}e^{-iK\left(t_{i}\right)}\,,
\end{equation}
and it can be shown that, at lowest order, the effective Hamiltonian
can be written as \cite{goldman2014}:
\begin{equation}
H_{\mathrm{eff}}=H_{0}+\tau\underset{n}{\sum}\frac{1}{n}\left[V_{n+},V_{n-}\right]+{\cal O}\left(\tau^{2}\right)\label{eq:Heff_oscilator}.
\end{equation}
This expansion relies on the small parameter $\tilde{V}\tau$ where $\tilde{V}$ is the typical energy scale of $V(t)$.
This expansion turns out to be very useful in the effective description
of ultracold atomic systems though care should be taken, in a case
by case scenario, in order to be sure about the convergence of the
series.

\subsubsection{Lattice Shaking}

The lattice shaking approach consists on having an external time dependent
optical potential that is changing in time in accordance to the previous
description. Then a change of basis is performed for a co-moving frame that,
along with a time average, will create an effective Hamiltonian with
the desired complex hopping. As an example, a brief
prescription is presented along the lines of the realization
in a $\mathrm{Rb}$ Bose-Einstein condensate \cite{Struck2012}. The Hamiltonian
considered is the usual tight-biding Hamiltonian in 2D with the usual
hopping and on-site part $H_{os}$ (by on-site it is intended one body
potential and scattering terms that act in single sites). In addition, there is
an extra time dependent potential:
\begin{equation}
H=-\underset{\vec{r},j}{\sum}t_{\vec{r}j}\hat{a}_{\vec{r}+\hat{j}}^{\dagger}\hat{a}_{\vec{r}}+H_{\mathrm{os}}+\underset{\vec{r}}{\sum}v_{\vec{r}}\left(t\right)\hat{a}_{\vec{r}}^{\dagger}\hat{a}_{\vec{r}}.
\end{equation}
The function $v_{\vec{r}}\left(t\right)$ is periodic on time
with period $\tau$: $v_{\vec{r}}\left(t\right)=v_{\vec{r}}\left(t+\tau\right)$. A unitary transformation on the states is performed and plugged in on the
Schr\"{o}dinger equation, thus defining new states $\left|\psi^{\prime}\right\rangle $
such that $\left|\psi\left(t\right)\right\rangle =U\left(t\right)\left|\psi^{\prime}\left(t\right)\right\rangle $.
The Hamiltonian becomes $H^{\prime}\left(t\right)=U\left(t\right)^{\dagger}HU\left(t\right)-iU\left(t\right)^{\dagger}\dot{U}\left(t\right)$
(where the dot stands for time derivative). The transformation is
given by
\begin{equation}
U\left(t\right)=e^{-i\int_0^t dt^{\prime} \sum_{\vec{r}} v_{\vec{r}}\left(t^{\prime}\right)\hat{a}_{\vec{r}}^{\dagger}\hat{a}_{\vec{r}}}\,.
\end{equation}
It is straightforward to see that this transformation cancels
the part of $H$ (which will be present also on $U^{\dagger}HU$)
corresponding to $v_{i}\left(t\right)\hat{a}_{\vec{r}}^{\dagger}\hat{a}_{\vec{r}}$.
On the other side, since this does not commute with the kinetic term,
a time dependence will be inherited by the hopping terms. For a set
of rapidly oscillating function $v_{i}\left(t\right)$ the Hamiltonian can
be replaced by an effective one, resulting from time averaging
over a period. The new hopping parameters will read:
\begin{equation}
t_{\vec{r}j}\rightarrow t_{\vec{r}j}\left\langle e^{i\Delta v_{\vec{r}j}}\right\rangle _\tau
\end{equation}
 where $\left\langle \right\rangle _{\tau}$ stands for the average over
a period: $\tau^{-1}\int_{0}^{\tau}dt$ and $\Delta v_{\vec{r}j}=v_{\vec{r}}\left(t\right)-v_{\vec{r+}\hat{j}}\left(t\right)-\left\langle v_{\vec{r}}\left(t\right)-v_{\vec{r+}\hat{j}}\left(t\right)\right\rangle_\tau $. 

\subsubsection{Laser-assisted hopping}

In this case the effective dynamics is induced by the coupling of
the atoms on the optical lattice with a pair of Raman lasers. A fundamental
ingredient consists on introducing an energy offset $\Delta$ on neighboring
sites. It is enough to consider such scenario along a single direction.
Considering a $2D$ lattice:
\begin{equation}
H=-t\underset{\vec{r},j}{\sum}\left(\hat{c}_{\vec{r}+\hat{j}}^{\dagger}\hat{c}_{\vec{r}}+\mathrm{h.c.}\right)+\frac{\Delta}{2}\underset{\vec{r}}{\sum}\left(-1\right)^{x}c_{\vec{r}}^{\dagger}\hat{c}_{\vec{r}}+V\left(t\right)
\end{equation}
where $\vec{r}=\left(x,y\right)$ runs through the lattice sites.
The offset term characterized by $\Delta$ can be obtained by tilting
the lattice, introducing magnetic gradients or through superlattices.
The potential $V\left(t\right)$ is the result of the two external
lasers that induce an electric field $E_{1}\cos\left(\vec{k}_{1}\cdot\vec{r}_{1}-\omega_{1}t\right)+E_{2}\cos\left(\vec{k}_{2}\cdot\vec{r}_{2}-\omega_{2}t\right)$.
It is assumed that the frequencies are fine-tuned such that they match
the offset $\omega_{1}-\omega_{2}=\Delta$. Neglecting fast oscillating
terms the potential is written as:
\begin{equation}
V\left(t\right)=2E_{1}E_{2}\underset{\vec{r}}{\sum}e^{i\left(\vec{k}_{R}\cdot\vec{r}-\Delta t\right)}c_{\vec{r}}^{\dagger}c_{\vec{r}}+\mathrm{h.c.}
\end{equation}
with $\vec{k}_{R}=\vec{k}_{1}-\vec{k}_{2}$. Then one can get the
effective Hamiltonian in two steps. First performing an unitary transformation
$\exp[{-it\frac{\Delta}{2}\sum_{\vec{r}}\left(-1\right)^{x}c_{\vec{r}}^{\dagger}\hat{c}_{\vec{r}}}]$
will create oscillatory hopping terms (with $\exp\left({\pm i\Delta t}\right)$ in
front). Then one may apply the previous formalism building an effective
Hamiltonian using Equation \ref{eq:Heff_oscilator}:
\begin{equation}
\begin{split}
H=&-t\underset{x,y}{\sum}\left(\hat{c}_{x,y}^{\dagger}\hat{c}_{x,y+1}+\mathrm{h.c.}\right)\\
&-\frac{2tE_{1}E_{2}}{\Delta}\underset{x\ \mathrm{even},y}{\sum}\left[\left(e^{i\vec{k}_{R}\cdot\vec{r}}-1\right)\left(e^{i\vec{k}_{R}\cdot\vec{r}}\hat{c}_{x,y}^{\dagger}\hat{c}_{x+1,y}+e^{-i\vec{k}_{R}\cdot\vec{r}}\hat{c}_{x-1,y}^{\dagger}\hat{c}_{x,y}\right)+\mathrm{h.c.}\right]+{\cal O}\left(\Delta^{-2}\right)
\end{split}
\end{equation}
It is clear that this generates complex hopping and looking more carefully
one finds that the lattice has a staggered flux. With a choice $\vec{k}_{R}=\left(\Phi,\Phi\right)$
(as also made in the experiment \cite{Aidelsburger2011}) one
can write upon a gauge transformation:
\begin{equation}
\begin{split}
H=&-t\underset{x,y}{\sum}\left(\hat{c}_{x,y}^{\dagger}\hat{c}_{x,y+1}+\mathrm{h.c.}\right)\\
&-\frac{2tE_{1}E_{2}\sin\Phi/2}{\Delta}\underset{x\ \mathrm{even},y}{\sum}\left[\left(e^{i\Phi y}\hat{c}_{x,y}^{\dagger}\hat{c}_{x+1,y}+e^{-i\Phi y}\hat{c}_{x-1,y}^{\dagger}\hat{c}_{x,y}\right)+\mathrm{h.c.}\right]+{\cal O}\left(\Delta^{-2}\right)
\end{split}
\end{equation}
where it is clear that a sequence of fluxes $\pm\Phi$ alternates in the plaquettes
along the $x$ direction. More refined techniques allow for the realization of systems with uniform fluxes \cite{Goldman2015}. In such systems the Chern number of the Hofstadter
bands was measured in \cite{Aidelsburger2014}.
It is worth noting that other kind of one-body terms, beyond the staggered
term, can be used as it was done in the first quantum simulations
of this model with ultracold atoms \cite{Aidelsburger2013,Miyake2013}.
In that case a linear potential is used. These kind of approaches
can be adapted to more general scenarios including different geometries
and multi-component species. The latter, for example, can be achieved
by introducing spin dependent potentials as done in \cite{Aidelsburger2013}.

\section{Simulation of gauge fields\label{sec:Simulating-gauge-fields}}

In the context of Abelian gauge theories, the goal of simulating gauge
fields consists in attributing dynamics to the complex phases on the
hopping parameters that were identified in the previous Section. In
order to construct such dynamics one should identify degrees of freedom
that will play the role of the gauge field. Several proposals have
been put forward which map the gauge degrees of freedom into some other
controllable variables. The platforms used include ultracold atoms,
trapped ions and superconducting qubits. They may be analogue or digital
quantum simulators and include Abelian or non-Abelian symmetries \cite{Weimer2010,glaetzle2014,tagliacozzo2013Abelian,Banerjee2012,notarnicola2015,kasper2016,kapit2011,zohar2013_2d+1,bazavov2015,Zohar2012,tewari2006,tagliacozzo2013,banerjee2013,stannigel2014,Zohar2013SU(2),Zohar2013,hauke2013,marcos2013,marcos2014,mezzacapo2015,doucot2003,laflamme2015,laflamme2016,Zohar2017,Zohar2017digitalgauge,Brennen2016,dehkharghani2017}.
A more detailed description of two particular approaches in analogue
cold atomic simulators will follow: the gauge invariance will be obtained by either penalizing with a large energy cost the non-physical states or by exploiting microscopic symmetries. The symmetries addressed
will be $U\left(N\right)$ and $SU(N)$. 

There are other symmetries which have been explored, namely $\mathbb{Z}_{n}$
\cite{Zohar2013,notarnicola2015} which, in particular, can provide
an alternative route towards $U\left(1\right)$ symmetry in the large
$n$ limit \cite{notarnicola2015} and can be addressed with similar
approaches. Proposal for the realization of $\mathbb{C}P\left(N-1\right)$
\cite{adda19781,eichenherr1978} models have been put forward in \cite{laflamme2015,laflamme2016}.
These models can serve as toy models for QCD and are also relevant
in studying the approach to the continuum limit, in the context of
D-theories, where the continuum limit is taken via dimensional reduction
\cite{brower1999,brower2004}. Furthermore other formulations are
possible for specific groups \cite{Tagliacozzo2014,mathur2005,anishetty2009,ZoharBurrello2015,ZoharBurrello2016}.
Gauge theories with Higgs fields have also been the target of quantum
simulation proposals \cite{Kuno2016,Kasamatsu2013,Kuno2017,GonzlezCuadra2017}.

Another relevant approach is the so-called quantum Zeno dynamics which
takes inspiration on the quantum Zeno effect, stating that
a system being continuously observed does not evolve on time.
Furthermore, if the measurement commutes with a certain part of the
Hamiltonian, then it can freeze a certain part of the Hilbert space
but still enables the dynamics in another subspace \cite{facchi2002}.
This feature can be used in order to freeze gauge dependent quantities
and let the system evolve in the gauge invariant subspace. The Hamiltonian
to be implemented has the form $H_{\mathrm{noise}}=H_{0}+H_{1}+\sqrt{2\kappa}\sum_{x,a}G_{x}^{a}$
where $H_{0}$ and $H_{1}$ are time independent and are, respectively,
gauge invariant and gauge variant parts of the Hamiltonian. The operators
$G_{x}^{a}$ are associated to the constraint one wishes to impose
$G_{x}^{a}\left|\psi\right\rangle =0$. In the case of gauge theories
$G_{x}^{a}$ are the generators of gauge transformations. An advantage
of this approach, with respect to the energy punishment approach of
the next Section, is that only linear terms on the generators must
be imposed on the Hamiltonian (energy punishment requires quadratic
terms). By other side leakage from the gauge invariant subspace of
the Hilbert space happens as a function of time, which does not happen
in the energy penalty approach. This approach was developed in \cite{stannigel2014}.

Another approach, that was successfully implemented in the first
quantum simulator of a gauge theory using trapped ions \cite{martinez2016},
is the digital quantum simulator \cite{lloyd1996}. The key idea consists
on in dividing the full time evolution operator $e^{-iHt}$ into smaller
pieces of sizes $\tau=t/N$ and apply time evolution of smaller parts
of the Hamiltonian at a time. Consider for example an Hamiltonian
which is a sum of $M$ contributions : $H=\sum_{\alpha}^{M}H_{\alpha}$.
Each part $H_{\alpha}$ can represent, for example, a nearest neighbor
spin interaction in which case only two spins are coupled on each
$H_{\alpha}.$ For large enough $N$ one can write:
\begin{equation}
e^{-iHt}=\left(e^{-iH\tau}\right)^{N}\simeq\left(\underset{\alpha=1}{\overset{M}{\prod}}e^{-iH_{\alpha}\tau}\right)^{N}
\end{equation}
Each time step can now be interpreted as an individual gate. While
in the analogue simulation the great difficult lies on building the
appropriate gauge invariant Hamiltonian, in digital quantum simulations
that is not a problem. The difficulty lies, however, in building an
efficient sequence of gates. Other then the scheme used in the first experimental
realization \cite{Muschik2017}, other proposals towards digital quantum simulations
of lattice gauge theories have been put forward \cite{tagliacozzo2013Abelian,Weimer2010,tagliacozzo2013,mezzacapo2015,Zohar2017,Zohar2017digitalgauge} 

\subsection{Gauge invariance from energy punishment\label{subsec:EnergyPunishement}}

The energy punishment approach is 
a quite general approach which allows for
the theoretical construction of models that will exhibit a given
symmetry in its low energy sector. It consists on building a Hamiltonian
which does not prohibit the symmetry violation to occur but instead
punishes it with a large energy. In a more concrete way, let suppose one
wants to implement a set of symmetries corresponding to a set
of generators $\left\{ G_{x}\right\} $ commuting with each other $\left[G_{x},G_{y}\right]=0$.
Furthermore consider a typical Hamiltonian $H_{0}$ which does not
respect these symmetries. Then one constructs the following Hamiltonian:
\begin{equation}
H=H_{0}+\Gamma\underset{x}{\sum}G_{x}^{2}\label{eq:Hpenalty}
\end{equation}
where $\Gamma$ is a large energy scale, meaning much larger than
the energy scales involved in $H_{0}$. Since $G_{x}$ are Hermitian
$G_{x}^{2}$ have non-negative eigenvalues. One can choose the lowest
eigenvalue to be zero by an appropriate definition of $G_{x}$. Then,
at low energy $\left(\ll\Gamma\right)$, the states will respect approximately
the condition $G_{x}\left|\psi\right\rangle \simeq0$. If not, this
would give a state automatically in an energy scale $\sim\Gamma$.
It is then possible to construct an effective Hamiltonian, valid in
low energy, which will respect the symmetries generated by $\left\{ G_{x}\right\} $.
Let $G$ be the projector operator on the subspace of the total Hilbert
space obeying $G_{x}\left|\psi\right\rangle =0$ and let $P=1-G$.
Then the low energy Hamiltonian can be written as:
\begin{equation}
H_{\mathrm{eff}}=GH_{0}G-\frac{1}{\Gamma}GH_{0}P\frac{1}{\underset{x}{\sum}G_{x}^{2}}PH_{0}G+{\cal O}\left(\Gamma^{-2}\right)\label{eq:Heff_pert_theory}
\end{equation}
which fulfills the symmetries. Within this framework an effective Abelian
gauge theory can be constructed. In non-Abelian theories the generators
of the gauge transformation do not commute and this construction fails.
There are, of course, several possible drawbacks even on the theoretical
level. For example the Hamiltonian \ref{eq:Heff_pert_theory}, even
though gauge invariant, may contain unwanted interactions or miss some particular terms which are present on the
target system. 

In order to construct a quantum simulator the first task is naturally
to map the degrees of freedom of the target theory into the laboratory
controlled ones, in this case the atomic variables. The matter fields,
which are fermionic, will naturally be described by fermionic atomic
species. Regarding gauge fields, the target will be the quantum links
formulation discussed in Section \ref{sec:QL}.
Therefore the goal consists on building the quantum links satisfying
the algebra $\left[L_{\vec{r},i},U_{\vec{r}',j}\right]=\delta_{ij}\delta_{\vec{r}\vec{r}'}U_{\vec{r}',j}$
and $\left[U_{\vec{r},i},U_{\vec{r}',j}^{\dagger}\right]=\delta_{ij}\delta_{\vec{r}\vec{r}'}2L_{\vec{r},i}/S\left(S+1\right)$.

\begin{figure}
\begin{centering}
\includegraphics[scale=0.4]{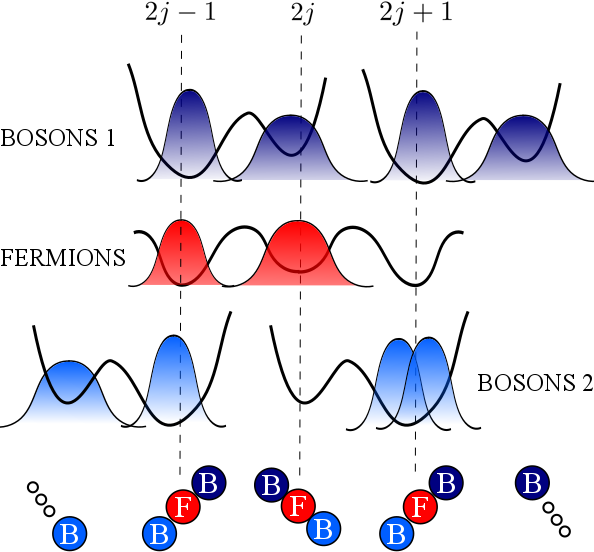}
\par\end{centering}
\caption{Superlattice configurations for the two boson species and the
fermionic one. Bosons of the species $1$ at an even site $2j$ can
only hop to $2j-1$ while a boson of species $2$ has only access
to the site $2j+1$. The Figure presents a an example of a gauge invariant
state configuration (on these three sites) where 
$G_{x}\left|\psi\right\rangle =0$.
\label{fig:Superlattice-punishement}}
\end{figure}

This can be achieved using the Schwinger representation. Given two
bosonic species $b^{\left(\sigma\right)}$ with $\sigma=1,2$ which
are associated to each link, one can write
\begin{equation}
U_{\vec{r}i}=\frac{1}{\sqrt{S\left(S+1\right)}}b_{\vec{r}i}^{\left(2\right)}{}^{\dagger}b_{\vec{r}i}^{\left(1\right)},\ L_{\vec{r}i}=\frac{1}{2}\left(b_{\vec{r}i}^{\left(2\right)}{}^{\dagger}b_{\vec{r}i}^{\left(2\right)}-b_{\vec{r}i}^{\left(1\right)}{}^{\dagger}b_{\vec{r}i}^{\left(1\right)}\right)\label{eq:Schwinger_bosons}
\end{equation}
Each link is loaded with a total of $2S$ bosons where $S$ is an
half integer. Then one has the desired representation for the quantum
links in terms of atomic variables. Now the variables are identified.
One then can then build a $d$ dimensional optical lattice where fermions
are allowed to hop among lattice points and in each links there are
a total of $2S$ bosons. For $1D$, the target Hamiltonian is of the
form:
\begin{equation} \label{hamstag}
H=-t\underset{n}{\sum}\left(c_{n}^{\dagger}U_{n}c_{n+1}+\mathrm{h.c.}\right)+m\underset{n}{\sum}\left(-1\right)^{n}c_{n}^{\dagger}c_{n}+\frac{g^{2}}{2}\underset{n}{\sum}L_{n}^{2}
\end{equation}
When comparing to the general structure of \ref{eq:General_Abelian_QL}
there are two differences: the plaquette term and the position-dependent coefficient of the kinetic term.
The plaquettes are naturally absent in $1D$, whereas the tunneling amplitude can be fixed by a gauge transformation $c_{n}\rightarrow\left(-i\right)^{n}c_{n}$.
The Hamilton \eqref{hamstag} has therefore the required structure and can be targeted with the Schwinger boson approach and it assumes the form:
\begin{equation}
H=-\frac{t}{\sqrt{S\left(S+1\right)}}\underset{n}{\sum}\left(c_{n}^{\dagger}b_{n}^{\left(\bar{\sigma}\right)}{}^{\dagger}b_{n}^{\left(\sigma\right)}c_{n}+\mathrm{h.c.}\right)+m\underset{n}{\sum}\left(-1\right)^{n}c_{n}^{\dagger}c_{n}+\frac{g^{2}}{8}\underset{n}{\sum}\left(b_{n}^{\left(2\right)}{}^{\dagger}b_{n}^{\left(2\right)}-b_{n}^{\left(1\right)}{}^{\dagger}b_{n}^{\left(1\right)}\right)^{2}\label{eq:H_U(1)_SchwingerBosons}
\end{equation}
The two last terms can be, in principle, implemented directly using
a proper tuning of the interactions between the bosons and the potential
for the fermions. The first term, instead, is a correlated hopping
between bosons and fermions which is obtained less easily. Furthermore the additional
terms that are not gauge invariant, like $b_{n}^{\left(\bar{\sigma}\right)}{}^{\dagger}b_{n}^{\left(\sigma\right)}$
and $c_{n}^{\dagger}c_{n+\hat{i}}$,
must be suppressed. This is solved by the energy punishment approach.
In general the non-gauge invariant Hamiltonian with the ingredients
described has the form:
\begin{equation}
\begin{split}
H_{0}=&-\underset{n,i}{\sum}\left[t_{F}\left(c_{n}^{\dagger}c_{n+1}+\mathrm{h.c.}\right)-t_{B}\left(b_{n}^{\left(2\right)}{}^{\dagger}b_{n}^{\left(1\right)}+\mathrm{h.c.}\right)\right]\\
&+\underset{n}{\sum}\left(v_{n}^{F}c_{n}^{\dagger}c_{n}+\underset{\sigma}{\sum}v_{n}^{B\sigma}b_{ni}^{\left(\sigma\right)}{}^{\dagger}b_{ni}^{\left(\sigma\right)}\right)+U\underset{n}{\sum}\left(b_{n}^{\left(2\right)}{}^{\dagger}b_{n}^{\left(2\right)}-b_{n}^{\left(1\right)}{}^{\dagger}b_{n}^{\left(1\right)}\right)^{2}\label{eq:H0_energypunishment}
\end{split}
\end{equation}
Using the generators for the $U\left(1\right)$ gauge symmetry
in Equation \ref{eq:u1-generators} one considers the full
Hamiltonian:
\begin{equation}
H=H_{0}+\Gamma\underset{n}{\sum}\left(L_{n}-L_{n-1}-c_{n}^{\dagger}c_{n}+\frac{1-\left(-1\right)^{n}}{2}\right)^{2}
\end{equation}
It is crucial that one has access to the interactions
that are introduced on the last term corresponding to the energy punishment. To see that this is the case it useful to be more specific about
the labels $\sigma$. One can take, as in \cite{Banerjee2012}, the
labels $\sigma=1,2$ meaning respectively left and right part of the
link, which can be thought to coincide with the lattice site. In this
way $b_{n}^{\left(2\right)}{}^{\dagger}b_{n}^{\left(1\right)}$ are
just regular hopping terms. Furthermore it is recalled that the total
number of bosons associated to each link is conserved. Therefore one
can write: $L_{n}=-S+b_{n}^{\left(2\right)}{}^{\dagger}b_{n}^{\left(2\right)}=S-b_{n}^{\left(1\right)}{}^{\dagger}b_{n}^{\left(1\right)}$.
This means that terms like $L_{n}^{2}$ and and $L_{n}L_{n-1}$ can
be written as a density-density interaction. Regarding the last case,
recall that $b_{n}^{\left(1\right)}$ and $b_{n-1}^{\left(2\right)}$
are effectively in the same site, see Figure \ref{fig:Superlattice-punishement}.
Now Equation \ref{eq:Heff_pert_theory} can be applied. The number
of particles in each site is a good quantum number to describe the
eigenstates of $G_{x}$. The number of particles in the site $j$
are denoted by $n_{j}^{F}=c_{j}^{\dagger}c_{j}$, $n_{j}^{1}=b_{j}^{\left(1\right)}{}^{\dagger}b_{j}^{\left(1\right)}$
and $n_{j}^{2}=b_{j-1}^{\left(2\right)}{}^{\dagger}b_{j-1}^{\left(2\right)}$.
The subspace of gauge invariant states is then characterized by:
\begin{equation}
n_{j}^{F}+n_{j}^{1}+n_{j}^{2}=2S+\frac{1-\left(-1\right)^{j}}{2}
\end{equation}

In the lowest order only the two last terms of \ref{eq:H0_energypunishment}
survive as any single hopping destroys the above relation. At the
next order there are three possible virtual processes that preserve
this condition. Up to some linear terms on the particle density operator,
they are:
\begin{enumerate}[i]
\item \label{cond1} Boson-boson hopping: a boson hops to the neighboring site on the same link
and another boson hops back. Gives rise to a boson density-density
interaction.
\item \label{cond2} Fermion-Fermion hopping: a fermion hops to a neighboring site and
then hops back. Only possible if neighboring site is unoccupied and
gives rise to a nearest neighbor fermion density-density interaction.
\item \label{cond3} Boson-Fermion hopping: a fermion hops to a neighboring site and a
boson belonging to the link that connects the two sites does the opposite
path. Gives rise to a correlated hopping.
\end{enumerate}
The terms coming from \ref{cond1} should be joined with the last term of \ref{eq:H0_energypunishment}
in order to form the correct kinetic term for the gauge fields. The terms in \ref{cond2} are somehow unwanted and correspond to a repulsion between
neighbor fermions $n_{j}^{F}n_{j+1}^{F}$. Naturally, they do not
spoil gauge invariance and their inclusion should not be a problem
\cite{Banerjee2012}. Finally the terms originating from \ref{cond3} give rise
to the correlated hopping responsible for the matter-gauge coupling
as written on the first term of \ref{eq:Schwinger_bosons}. There
is another issue which should be addressed. From the beginning it
was assumed that the the number of bosons in each link is conserved.
In particular this means that bosons are not allowed to pass to a
neighboring link. In order to guarantee this condition in an experiment
one should introduce an extra bosonic species and this is the reason
that bosons in neighboring links were represented with different colors
on Figure \ref{fig:Superlattice-punishement}. Then one bosonic species
is trapped on the even links and the other in the odd links. This
will prevent bosonic hopping between links. A numerical study of real
dynamics of the the model as well as accuracy of the effective gauge
invariance obtained was also done in \cite{Banerjee2012}.

Finally, in a possible experimental realization, the first fundamental
step is to guarantee that the system is initialized on a gauge invariant
state. This can be done by loading the atoms in a deep lattice such
that they are in Mott phase. Afterwards the system should evolve according
to the fine tuned Hamiltonian described above (after lowering the
lattice barriers). Finally measures of relevant quantities can be
performed.

This principle is valid in higher dimensionality where one has
to face the difficulty of generating plaquette terms. This was done
for the pure gauge in \cite{Zohar2011} and \cite{Zohar2012} by suitably allowing hopping between links. In the first case each link has an infinite dimensional Hilbert space that is represented by a Bose-Einstein condensate. In
the second the proposal is simplified by considering a quantum link
model.

\subsection{Gauge invariance from many body interaction symmetries}

This approach consists on building a lattice which will have the necessary
local gauge invariance arising from microscopic symmetries. Specific
proposals may vary significantly even though the same principle is used. For example in \cite{dehkharghani2017} the simulation
is built upon the global symmetry conserving the total number of excitations
and is achieved via a state-dependent hopping. In turn, see for example
\cite{Zohar2013,Kasper2017}, are built upon conservation of angular
momentum. For concreteness the later approach will be described in
more detail below. In the case of \cite{banerjee2013} $SU\left(N\right)$
symmetries of the ground state manifold of alkaline-earth-like atoms
could be exploited in order to built non-Abelian gauge theories. 

Symmetries only allow for certain type of processes to occur and, by exploiting
these constraints, one can build a gauge symmetry. This can be done,
as said before, considering angular momentum conservation. 
The Schwinger model is taken as an illustrative example. Bosons,
that will make up the gauge fields, are placed at the two boundaries of the links.
Because the goal consists, partially, in forbidding gauge dependent
terms like simple boson or fermion hopping, the lattice should be spin
dependent. In this way a single hopping is forbidden as it does not
conserve angular momentum. By other side one should guarantee that
correlated spin between bosons and fermions is allowed. This can be
achieved by a judicious choice of respective hyperfine angular momentum
in each lattice site. For concreteness, consider a single link connecting
two sites and a total of two bosonic ($b^{\left(1\right)}$,$b^{\left(2\right)}$)
and two fermionic species ($c$,$d$). The site at the left of the
link can only be populated by $c$ while the right side by $d$. Analogously
the left end of the link can only be populated by $b^{\left(1\right)}$
while the right end can only be populated by $b^{\left(2\right)}$.
Then the conditions described above for allowed/forbidden hopping
are automatically satisfied if one chooses the hyperfine angular momentum
of each atomic species to satisfy:
\begin{equation}
m_{F}\left(d\right)-m_{F}\left(c\right)=m_{F}\left(b^{\left(1\right)}\right)-m_{F}\left(b^{\left(2\right)}\right)
\end{equation}
It is intended that the lattice is, indeed, spin dependent so that $m_{F}\left(d\right)\neq m_{F}\left(c\right)$
and $m_{F}\left(b^{\left(1\right)}\right)\neq m_{F}\left(b^{\left(2\right)}\right)$.
In other words, what this means is that the difference
of angular momentum caused by a fermion hop can be exactly compensated
by a bosonic hop in the opposite direction. This leads directly to
the correlated hopping desired which, in fact, comes from the scattering terms between bosons and fermions.
The only other allowed scattering term between fermions and bosons
correspond to density-density interactions like $c^{\dagger}c\left(b^{\left(2\right)}{}^{\dagger}b^{\left(2\right)}+b^{\left(1\right)}{}^{\dagger}b^{\left(1\right)}\right)$.
These are just linear terms on the fermionic number operator due to
the conservation of the total number of bosons per link. Summing over
all lattice sites will give just a constant shift of the energy.
The scattering terms between bosons give rise to the gauge kinetic
term as before (in $1+1$ dimensions).

Again, for higher dimensionality, there is a non-trivial extra step
consisting on building plaquette interactions. If plaquettes are ignored
and the model described above is loaded on an higher dimensional lattice
the result corresponds to the strong coupling limit of the gauge theory. 

\begin{figure}
\begin{centering}
\includegraphics[scale=0.75]{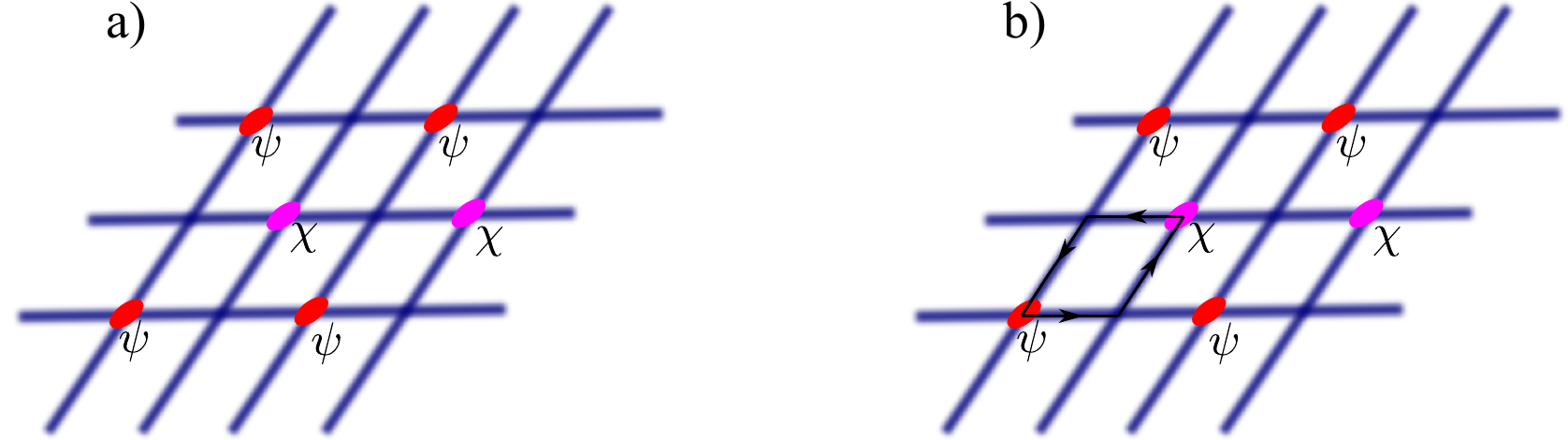}
\par\end{centering}
\caption{Loop method for obtaining the plaquette terms. In the panel a) it is depicted the positions of the auxiliary fermions
that are used to construct the plaquette term using gauge invariant
building blocks. One of the species, say $\psi$, is represented in
red and placed on sites with both coordinates even. In turn $\chi$,
in pink, is placed on sites with both coordinates odd. This correspond
to the ground-state of \ref{eq:H_loop}. In the panel b) it is represented
a virtual process that gives rise to a plaquette term\label{fig:loop_method}.}
\end{figure}

The plaquette terms can be achieved
by the so-called loop method. It uses perturbation theory in a similar
way that was used in the energy penalty approach. 
In order to discuss the essence of the construction of the plaquette terms, one can consider just the pure gauge theory. The target Hamiltonian is 
\begin{equation}
H_\mathrm{target}=\frac{g^{2}}{2}\underset{\vec{r},i}{\sum}L_{\vec{r}i}^{2}-\frac{1}{4g^{2}}\underset{\square}{\sum}\left(U_{\square}+U_{\square}^{\dagger}\right).
\end{equation}
The description will be specialized for $2+1$ dimensions but the
theoretical construction for higher dimensions is analogous. The construction of the plaquette term relies on a perturbative expansion similar to \ref{eq:Hpenalty} but, in this case, $H_{0}$
is already a gauge invariant Hamiltonian. For reasons that will be
explained below one should have two fermionic species, say $\chi$
and $\psi$, and build the trivial part of the generalization of the
$1+1$ process:
\begin{equation}
H_{0}=-t\underset{\vec{r},i}{\sum}\left(\psi_{\vec{r}}^{\dagger}U_{\vec{r}i}\psi_{\vec{r}+\hat{i}}+\chi_{\vec{r}}^{\dagger}U_{\vec{r}i}\chi_{\vec{r}+\hat{i}}+\mathrm{h.c.}\right)+\frac{g^{2}}{2}\underset{\vec{r},i}{\sum}L_{\vec{r}i}^{2}\,.\label{eq:H0_LoopMethod}
\end{equation}
The fermionic species are auxiliary and in the effective model they
will be integrated out. There should be no interacting term between
them. Here the energy penalty must enforce
the following conditions at each site $\vec{r}=\left(r_{1},r_{2}\right)$:
\begin{itemize}
\item there is a fermion $\psi$ if both $r_{1}$ and $r_{2}$ are even
\item there is a fermion $\chi$ if both $r_{1}$ and $r_{2}$ are odd
\item no fermion otherwise
\end{itemize}
The positions of these fermions is represented on Figure \ref{fig:loop_method}
a). This kind of constraint can be obtained, for large $\Gamma$,
with a Hamiltonian of the form:
\begin{equation}
H_\mathrm{penalty}=-\Gamma\underset{\vec{r}}{\sum}\left[\frac{\left(1+\left(-1\right)^{r_{1}}\right)\left(1+\left(-1\right)^{r_{2}}\right)}{4}\psi_{\vec{r}}^{\dagger}\psi_{\vec{r}}+\frac{\left(1-\left(-1\right)^{r_{1}}\right)\left(1-\left(-1\right)^{r_{2}}\right)}{4}\chi_{\vec{r}}^{\dagger}\chi_{\vec{r}}\right]\label{eq:H_loop}
\end{equation}
Through perturbation theory, according to \ref{eq:Heff_pert_theory}, one gets the plaquette terms at fourth order. This process is "cleaner" if the $U_{\vec{r}}$ in \ref{eq:H0_LoopMethod} are considered unitary. IN particular we may consider a unitary limit,in which the total spin  of the quantum link goes to infinity: $S\rightarrow =+\infty$. Order by order:
\begin{enumerate}
\item Only the pure gauge part of \ref{eq:H0_LoopMethod} contributes, no
fermionic term occurs.
\item Trivial constant contribution assuming that $U_{n}$ are unitary.
The virtual process giving rise to this contribution is a single link
interaction where a fermionic-bosonic correlated hopping occurs back
and forth restoring the initial state. There are never fermions on
the neighbor lattice site. In turn in the unitary limit there is
an infinite number of bosons such that $\left[U,U^{\dagger}\right] \to 0$.
In the case of finite bosonic number, extra contribution corresponding to
a renormalization of the pure gauge term of \ref{eq:H0_LoopMethod} will appear, together with another term which can be discarded by application of the Gauss law.
\item Trivial constant contribution assuming that $U_{n}$ are unitary.
Virtual contributions evolving links constitute again back and forth
hopping plus a pure gauge term at any stage of the process. The
extra contributions coming from considering a finite number of boson
per link cannot be disregarded trivially as second order for
this case.
\item Gives the desired plaquette term plus renormalization of the pure
gauge term of \ref{eq:H0_LoopMethod} assuming that $U_{n}$ are unitary.
The last case corresponds to the virtual process where a fermion goes
around a plaquette and returns to the initial place. This virtual
process is represented on Figure \ref{fig:loop_method} b). Naturally,
in the non-unitary case, more terms appear.
\end{enumerate}

Plaquette terms only appear at fourth order. However, in the unitary limit, most contributions are trivial. One can then see that it is effectively a second order contribution \cite{zohar2015}. 

When one considers a finite number of bosons in the links there are
extra contributions appearing which cannot be disregarded. As in the
case of the energy penalty, these contributions, even though unwanted,
can be tolerated as they are naturally gauge invariant. However one
should guarantee that these extra contributions are not more important
than the plaquette term which is the target term. That can be achieved
if the coupling term is parameterized is $g^{2}$ is taken to be small
in units of $t$. By taking $g^{2}\sim t^{2}/\Gamma$ one makes the
unwanted terms at third order effectively of the same order as the
plaquettes and unwanted terms of the fourth order effectively of higher
order than the plaquettes.

On top of these, an extra species
of fermions can be introduced to play the role of matter fields. They will consist,
in the initial Hamiltonian, to the usual correlated hopping with the bosons. Furthermore the staggered mass term (of Equation \ref{eq:H_KG_general_d})
should also be introduced. In the unitary case this extra piece commutes
with the interacting part of \ref{eq:H0_LoopMethod} and no further
contribution is obtained in perturbation theory. In the truncated
case there is an extra (gauge invariant) correlated hopping coming
at third order. Another different aspect of the introduction of dynamical
fermions is that the Gauss law ($\sum_{i}L_{\vec{r}i}-L_{\vec{r}-\hat{i},i}=\mathrm{const}$)
can no longer be used to trivialize terms. The divergence of the electric gets a contribution from the charge density of the dynamical fermions. Nonetheless it can still be employed and the extra charge
density terms can be compensated on the initial Hamiltonian if proper
fine tuning is available experimentally.

In \cite{Kasper2017} it was proposed a realization of the Schwinger
($1+1$) model using a mixture of $^{23}\mathrm{Na}$ for the bosons
and $^{6}\mathrm{Li}$ for the fermions as well as an extensive study
on the influence of the finiteness of the number of bosons per link
in that case.

\subsection{Encoding in $1+1$ fermions \label{subsec:Encoding-in-}}

The case of the Schwinger model, $1+1$ Dirac fermions coupled to
a gauge field, is an interesting experimental and theoretical playground.
It shares some non-trivial features with QCD like confinement, chiral
symmetry breaking and a topological theta vacuum \cite{wiese2013}.
However, due to its simplicity, it allows for analytical and numerical
studies which may become significantly harder in more complicated
theories. Furthermore it was the target of the first experimental
implementation of a lattice gauge theory \cite{martinez2016}. In
the context of quantum simulations it may not only provide the entrance
door towards more complicated experimental realizations but also a
way of benchmarking experimental techniques.

One of the reasons why this model bares an intrinsic simplicity, as
mentioned previously, is the fact that the gauge fields are non-dynamical.
This is reflected on the absence of plaquette terms in the Hamiltonian
formulation. Furthermore the Gauss law fixes the gauge field and can be used to integrate out its degrees of freedom. This results in
a long-range interacting model which will be addressed next. In the following the lattice Hamiltonian formulation is considered for $N$ lattice sites:
\begin{equation}
H=-it\overset{N-1}{\underset{n=1}{\sum}}\left[c_{n}^{\dagger}U_{n}c_{n+1}-\mathrm{h.c}\right]+m\overset{N}{\underset{n=1}{\sum}}\left(-1\right)^{n}c_{n}^{\dagger}c_{n}+\frac{g^{2}}{2}\overset{N-1}{\underset{n=1}{\sum}}L_{n}^{2}\label{eq:PBC_KG_Schwinger}
\end{equation}
Here an infinite dimensional Hilbert space per link is considered,
therefore the operators $U_{n}$ are unitary and the non-trivial commutation relations
on the links are given by $\left[L_{m},U_{n}\right]=U_{n}\delta_{mn}$.
Equivalently the link can be written as $U_{n}=e^{i\theta_{n}}$.
The Gauss law is imposed in accordance with the relations \eqref{gausslaw} and \eqref{eq:u1-generators}.
This model can be formulated in terms of Pauli spin operators \cite{Banks1976}
through the Jordan-Wigner transformation:
\begin{equation}
\left\{ \begin{array}{c}
c_{n}=\underset{l<n}{\prod}\left(i\sigma_{z}\left(l\right)\right)\sigma^{-}\left(n\right)\\
c_{n}^{\dagger}=\underset{l<n}{\prod}\left(-i\sigma_{z}\left(l\right)\right)\sigma^{+}\left(n\right)
\end{array}\right.\label{eq:Jordan_Wigner}
\end{equation}
where $\sigma_{i}\left(l\right)$ represent the Pauli matrices in the site $l$ and $\sigma^{\pm}\left(n\right)=\sigma_{x}\left(n\right)\pm i\sigma_{y}\left(n\right)$.
In terms of the spins the Gauss law is determined by:
\begin{equation}
G_{n}=L_{n}-L_{n-1}-\frac{1}{2}\left(\sigma_{z}\left(n\right)+\left(-1\right)^{n}\right)\,.
\end{equation}
By restricting ourselves to the physical space, through the Gauss law $G_{n}\left|\psi\right\rangle =0$, the link variables can be almost completely eliminated. Using periodic boundary
conditions ($L_{0}=L_{N}$) one finds:
\begin{equation}
L_{n}=L_{0}+\frac{1}{2}\underset{l=1}{\overset{n}{\sum}}\left(\sigma_{z}\left(l\right)+\left(-1\right)^{n}\right)\,.
\end{equation}
The value of $L_{0}$ is a parameter of the theory and corresponds
to a background field. For simplicity it will be taken to zero at
the present discussion. By using the above relations the Hamiltonian \ref{eq:PBC_KG_Schwinger} can be rewritten as:
\begin{equation}
H=t\overset{N}{\underset{n=1}{\sum}}\left[\sigma^{+}\left(n\right)e^{i\theta_{n}}\sigma^{-}\left(n+1\right)+\mathrm{h.c}\right]+\frac{m}{2}\overset{N}{\underset{n=1}{\sum}}\left(-1\right)^{n}\sigma_{z}\left(n\right)+\frac{g^{2}}{8}\overset{N}{\underset{n=1}{\sum}}\left[\underset{l=1}{\overset{n}{\sum}}\left(\sigma_{z}\left(l\right)+\left(-1\right)^{n}\right)\right]^{2}\label{eq:encoded_H}
\end{equation}
where a trivial constant term was dropped. The remaining gauge field
variable $\theta_{n}$ can be eliminated by a residual gauge transformation
\cite{Hamer1997}:
\begin{equation}
\sigma^{\pm}\left(n\right)\rightarrow\sigma^{\pm}\left(n\right)\underset{j<n}{\prod}e^{\pm i\theta_{j}}\,.
\end{equation}
This is a non-trivial transformation as $\theta$'s are operators. More precisely, the above relation should be seen as defining a new set of operators $\bar{\sigma}^{\pm}\left(n\right) =\sigma^{\pm}\left(n\right)\prod_{j<n}e^{\pm i\theta_{j}}$. The  $\bar{\sigma}$ still respect the angular momentum algebra between each other. Therefore they are still spin operators on the sites of the lattice, despite acting non-trivially on the links. Since the links degrees of freedom are being traced out using the Gauss law, one can arrive at an effective spin model for the sites.
Plugging this transformation and expanding the interaction term, the
resulting model is a long-range interacting spin model:
\begin{equation}
H=t\overset{N}{\underset{n=1}{\sum}}\left[\sigma^{+}\left(n\right)\sigma^{-}\left(n+1\right)+\mathrm{h.c}\right]+\overset{N}{\underset{n=1}{\sum}}\left(\frac{m}{2}\left(-1\right)^{n}-\frac{g^{2}}{8}\left(1-\left(-1\right)^{n}\right)\right)\sigma_{z}\left(n\right)+\frac{g^{2}}{4}\overset{N-2}{\underset{n=1}{\sum}}\underset{l=1}{\overset{N-1}{\sum}}\left(N-l\right)\sigma_{z}\left(n\right)\sigma_{z}\left(l\right)
\end{equation}

This is a useful formulation for quantum simulations since the total of $N$ particles
and $N-1$ gauge fields are simulated by just $N$ spins (with exotic
long-range interactions), thanks to the gauge invariance. The difficulty was moved towards an efficient
way of implementing the long-range asymmetric interaction between
spins. This Hamiltonian was implemented as a digital quantum simulator
in \cite{martinez2016} using trapped ions ($^{40}\mathrm{Ca}^{+}$).
The system was composed of four qubits. The Schwinger mechanism of
pair creation of particle-antiparticle was explored, as well as real
time evolution of entanglement in the system. Based on the staggering prescription  in Section \ref{sec:dynamical},
a particle on an odd site corresponds to the vacuum and
a hole as an antiparticle (the contrary holds for particles in the
even sites). Following this picture the number of particles at the
site $n$ is given by $\nu_{n}=\left(1-\left(-1\right)^{n}\right)/2+\left(-1\right)^{n}c_{n}^{\dagger}c_{n}$
and therefore a relevant observable is the particle density $\nu\left(t\right)=\left(2N\right)^{-1}\sum_{n}\left\langle 1+\left(-1\right)^{n}\sigma_{z}\left(n\right)\right\rangle $.
Starting from a bare vacuum ($\nu\left(0\right)=0$) it is observed
a rapid increase of the particle density followed by a decrease which
is due to particle/anti-particle recombination. Also the vacuum persistence $G\left(t\right)=\left\langle 0\right|e^{-iHt}\left|0\right\rangle $
and entanglement were evaluated. The latter is done by reconstructing
the density matrix and evaluating the entanglement in one half of
the system with the other half through logarithmic negativity. Entanglement
is produced through particle creation that get distributed across
the two halves. More detail on the simulation and experimental results
can be found in \cite{martinez2016,Muschik2017}. Future challenges
include the simulation of larger systems as well higher dimensionality
and non-Abelian symmetries.

\begin{acknowledgement}
The authors want to thank the organizers of the Natal's school: 
Pasquale Sodano, Alvaro Ferraz, Kumar S. Gupta and Gordon Semenoff. 
They are also pleased to thank the participants to the course, 
in particular T. J. G. Apollaro, V. E. Korepin, T. Macr\`i, 
G. Mussardo, E. Tonni and J. Viti, for useful 
discussions and to have contributed to set a stimulating and pleasant 
atmosphere. Finally, special thanks go to Marcello Dalmonte for discussions 
and common work on the topics discussed in this chapter.
\end{acknowledgement}
%

\bibliographystyle{unsrt}

\bibliography{biblio_PB_B_T}

\end{document}